\documentclass[pra,aps,twocolumn,showpacs]{revtex4}
\usepackage{amsmath,graphicx,epsfig}

\def\abr#1{{\langle #1 \rangle}}

\def\ket#1{{|#1\rangle}}

\def\ts#1{{_{\mbox{\scriptsize #1}}}}

\def\threej(#1,#2)(#3,#4)(#5,#6){\begin{pmatrix}#1&#3&#5\\#2&#4&#6\end{pmatrix}}
\def\sixj(#1,#2,#3)(#4,#5,#6){\begin{Bmatrix}#1&#2&#3\\#4&#5&#6\end{Bmatrix}}
\def\ninej(#1,#2,#3)(#4,#5,#6)(#7,#8,#9){\begin{Bmatrix}#1&#2&#3\\#4&#5&#6\\#7&#8&#9\end{Bmatrix}}

\newcommand{\ki}{\Gk_I}

\newcommand{\mc}{\mathcal}
\newcommand{\ms}[1]{\mspace{#1mu}}
\newcommand{\prn}[1]{\left(#1\right)}

\newcommand{\sbr}[1]{\left[#1\right]}

\newcommand{\Gd}{\delta}

\newcommand{\Gg}{\gamma}

\newcommand{\Gk}{\kappa}

\def\cg(#1,#2)(#3,#4)(#5,#6){{\langle#1,#2,#3,#4}\ket{#5,#6}}

\def\fig_width{3. in} 
\draft

\newlength{\defbaselineskip}
\newcommand{\setlinespacing}[1]%
           {\setlength{\baselineskip}{#1 \defbaselineskip}}
\newcommand{\doublespacing}{\setlength{\baselineskip}%
                           {2.0 \defbaselineskip}}

\begin{document}

\title{Relaxation of atomic polarization in paraffin-coated cesium vapor cells} 
\author{M. T. Graf}
\author{D. F. Kimball}
\author{S. M. Rochester}
\author{K. Kerner}
\author{C. Wong}
\author{D. Budker}\email{budker@socrates.berkeley.edu}
\affiliation{Department of Physics, University of California at
Berkeley, Berkeley, California 94720-7300}

\author{E. B. Alexandrov}
\author{M. V. Balabas}
\affiliation{S. I. Vavilov State Optical Institute, St.
Petersburg, 199034 Russia}


\author{V. V. Yashchuk}
\affiliation{Advanced Light Source Division, Lawrence Berkeley
National Laboratory, Berkeley, California 94720}


\date{\today}


\begin{abstract}
The relaxation of atomic polarization in buffer-gas-free,
paraffin-coated cesium vapor cells is studied using a variation on
Franzen's technique of ``relaxation in the dark'' [Franzen, Phys.
Rev. {\bf 115}, 850 (1959)].  In the present experiment,
narrow-band, circularly polarized pump light, resonant with the Cs
D2 transition, orients atoms along a longitudinal magnetic field,
and time-dependent optical rotation of linearly polarized probe
light is measured to determine the relaxation rates of the atomic
orientation of a particular hyperfine level. The change in
relaxation rates during light-induced atomic desorption (LIAD) is
studied. No significant change in the spin relaxation rate during
LIAD is found beyond that expected from the faster rate of
spin-exchange collisions due to the increase in Cs density.
\end{abstract}
\pacs{PACS. 32.80.Bx, 34.50.Dy, 79.20.La}




\maketitle

\doublespacing

\section{Introduction}

Coating the walls of an alkali-metal vapor cell with paraffin wax
reduces the relaxation rate of atomic polarization by up to four
orders of magnitude
\cite{Robinson,Bouchiat,Alexandrov1,Alexandrov2}.  Long-lived
atomic polarization (relaxation times of $\sim 1~{\rm{s}}$ has
been observed) enables extremely sensitive measurements of
magnetic fields
\cite{Claude,Alexandrov1,Alexandrov2,NMOR,Alexandrov3}, enhances
nonlinear optical effects at low light powers (see the review
\cite{NMOEreview} and references therein), and may make possible
precision tests of fundamental symmetries
\cite{UW_edm,Yashchuk,Kimball}. Paraffin-coated cells have drawn
attention in the study of light propagation dynamics
\cite{SlowLight,FastLight}, for the generation of spin-squeezed
states \cite{Bigelow}, and the creation and study of high-rank
polarization moments \cite{HexaDecapole}. There has also been
renewed interest in the application of paraffin-coated cells in
miniaturized atomic clocks \cite{NISTpaper}.  In spite of their
wide and varied application and several detailed studies of their
spin-relaxation properties
\cite{BouchiatPhD,Bouchiat,Gibbs,Liberman,BalabasRID,Vanier,AleksandrovK},
there is still much to learn about the mechanisms of spin
relaxation in paraffin-coated cells.

In this study, we use a variation on Franzen's classic technique
of ``relaxation in the dark'' \cite{FranzenRID} to elucidate the
mechanisms for spin-relaxation of cesium atoms in buffer-gas-free,
paraffin-coated cells (prepared in the manner described in
Ref.~\cite{ourLIAD}). A circularly polarized laser beam (the pump
beam), tuned to resonance with one of the hyperfine components of
the D2 transition, propagates along the direction of an applied
magnetic field ($\hat{z}$) and polarizes the Cs atoms. The pump
beam is abruptly blocked by a shutter and the decay of the atomic
polarization is monitored by measuring optical rotation of a weak,
linearly polarized probe beam (propagating collinearly with the
pump beam).

A key challenge in the interpretation of the measurements is to
determine the physical meaning of the optical signal, in
particular how it relates to the atomic polarization in the cell.
In the usual implementation of Franzen's technique, in which
transmission of a circularly polarized probe is measured to
determine the relaxation of atomic polarization, the observed
signal depends on two quantities \cite{Bouchiat,HapperReview}: the
longitudinal electronic polarization $\abr{S_z}$ and the
population difference between the two ground-state hyperfine
levels, proportional to $\abr{\bf S \cdot I}$ (here ${\bf S}$
represents the electron spin and ${\bf I}$ represents the nuclear
spin).  The relation of the optical signal to the relaxation of
$\abr{S_z}$ and $\abr{\bf S \cdot I}$ depends on the spectral
properties and polarization of the probe light \cite{Bouchiat}.

Furthermore, the quantities $\abr{S_z}$ and $\abr{\bf S \cdot I}$
can relax with several different time constants depending on the
relaxation mechanisms (e.g., spin-exchange collisions between Cs
atoms, electron-randomization collisions with the cell walls,
relaxation due to exchange of atoms between a metal sample in the
stem of the cell and the vapor phase in the volume of the cell --
known as the ``reservoir effect'' \cite{Bouchiat}). Consequently,
when narrow-band laser light is used, the observed signal in
Franzen's method relaxes with multiple rates that can be difficult
to distinguish.

In contrast (as discussed in Sec.
\ref{SubSec:PrincipleOfRIDOVOR}), by observing optical rotation of
a linearly polarized probe beam as we do in the present
experiment, under the magnetic field conditions of our experiments
(where for most measurements $B \lesssim 15~{\rm G}$) the observed
signal is well described with only two exponentials. Furthermore,
the amplitudes of the rotation associated with the two
exponentials turn out to be opposite in sign, allowing a clear
distinction between the associated rates. We analyze our
experimental results using the concepts of atomic polarization
moments (see, for example,
Refs.~\cite{OkunevichBook,NMOEreview,OurBook,Stenholm,Blum}), and
find that under appropriate conditions the optical signal is
sensitive only to the rank-one multipole moment (orientation).

We have used this technique of relaxation in the dark observed via
optical rotation to study the change in the relaxation rates of
atomic polarization when a paraffin-coated vapor cell is exposed
to non-resonant light that causes desorption of alkali atoms from
the paraffin coating (light-induced atomic desorption, LIAD, see
Ref.~\cite{ourLIAD} and references therein).  LIAD is of interest
as a method for rapid control of the vapor density in
paraffin-coated cells; LIAD can also be used as a technique for
the study of wall coatings.  We find no significant change in
spin-relaxation rates in the cell during LIAD beyond that expected
from the faster rate of spin-exchange collisions due to the
increase in Cs vapor density. This indicates that LIAD does not
significantly affect the relaxation properties of the coating.  In
contrast, when the alkali density is increased by heating the
cell, our work shows evidence of a significant increase in
relaxation caused by electron-randomization collisions.

\section{Relaxation in the dark observed via optical rotation}\label{Sec:RIDtechnique}

\subsection{Principle of measurement technique}\label{SubSec:PrincipleOfRIDOVOR}

When optical pumping with circularly polarized laser light is
performed, in general the populations of the ground state
hyperfine levels are altered and the atomic medium acquires both
orientation and alignment along the direction of light propagation
(see, for example, Refs.~\cite{NMOEreview,OurBook}). The
orientation and alignment of a collection of atoms can be
characterized using the density matrix formalism (as described,
for example, in
Refs.~\cite{NMOEreview,OurBook,Stenholm,Blum,HapperReview} -- also
see several recent articles focusing specifically on atomic
polarization in paraffin-coated cells
\cite{Okunevich1,Okunevich2,Okunevich3}). Orientation corresponds
to the rank $\kappa=1$ irreducible tensor component of the density
matrix and alignment corresponds to the rank $\kappa=2$ component,
while the population corresponds to the $\kappa=0$ component. For
a state with total angular momentum $F$, the multipole moments
$\rho_q^{(\kappa)}$ are related to the usual Zeeman components of
the density matrix $\rho_{M,M'}$ (where $M,M'$ refer to Zeeman
sublevels) via the equation \cite{Omont,Varshalovich}
\begin{align}
\rho_q^{\kappa}= \sum_{M,M'=-F}^{F}
(-1)^{F-M'}\cg(F,M)(F,-M')(\kappa,q) \rho_{M,M'}~,
\end{align}
where $\cg(F,M)(F,-M')(\kappa,q)$ is the appropriate Clebsch-Gordan coefficient. Note
that the atomic polarization moments may be of rank $\kappa = 0, 1, 2, ... , 2F$ for a
given hyperfine level.

If circularly polarized pump light propagates in the $\hat{z}$ direction (chosen here to
be the quantization axis), alkali atoms can acquire nonzero polarization moments with
$q=0$ in each hyperfine level. In our experiments, the creation of moments with $q \neq
0$ is further suppressed by the application of a longitudinal magnetic field, which
averages out any transverse polarization (to which the experiment might be sensitive due
to misalignment of the pump and probe beams). While moments higher than $\kappa=2$ can be
created by the pump light, because the light power in the probe beam is very low, we can
safely assume we are only probing the lowest rank polarization moments, $\kappa = 0,1,2$
(those for which the optical rotation is independent of probe light power
\cite{HexaDecapole}). Therefore in our analysis we consider only the moments
$\rho_0^{(0)}$ (population), $\rho_0^{(1)}$ (orientation along $z$), and $\rho_0^{(2)}$
(alignment along $z$) in each hyperfine level.

We assume that three different types of relaxation processes for
ground state atomic polarization are possible in the
paraffin-coated cell: (1) electron-randomization collisions with
the paraffin-coated cell wall or perhaps gaseous impurities, (2)
spin-exchange collisions between the Cs atoms, and (3) a process,
such as the reservoir effect \cite{Bouchiat}, which relaxes all
polarization moments at the same rate (denoted as {\emph{uniform
relaxation}}):
\begin{align}
\frac{d}{dt}\rho^\Gk_q(F) & = \\
&\sbr{\frac{d}{dt}\rho^\Gk_q(F)}_\text{ER} + \sbr{\frac{d}{dt}\rho^\Gk_q(F)}_\text{SE} +
\sbr{\frac{d}{dt}\rho^\Gk_q(F)}_\text{U}~. \nonumber
\end{align}
Estimates and experimental evidence (discussed in
Sec.~\ref{SubSec:AssumptionsVerified}) show that relaxation due to
magnetic field gradients can be neglected. The probe light power
used (3-5~${\rm \mu W}$) is sufficiently low that relaxation due
to optical pumping by the probe light can be neglected as well
(see Sec.~\ref{SubSec:AssumptionsVerified}).

Electron-randomization collisions completely randomize the polarization of the valence
electron, but the nuclear spin of the Cs atom is altered only due to the fact that
hyperfine interactions recouple the electron spin to the nuclear spin after the
collision. Thus the total atomic polarization takes many [about $(2I+1)^2$ for $I \gg 1$]
collisions to relax (this is known as the nuclear slow-down effect).  Taking these
effects into account, for a rate $\gamma_{er}$ of electron-randomization collisions, the
equation describing the relaxation of atomic polarization moments is (for $\kappa > 0$)
\cite{OkunevichBook}
\begin{align}
    \sbr{\frac{d}{dt}\rho^\Gk_q(F)}_\text{ER}
        =-\Gg_{er}
            \sum_{F_{\ms{-2}1}}
            \sum_{\ki}\mc{P}^{F}_{F_{\ms{-2}1}}\!\prn{\ki,\Gk}
            \rho^{\Gk}_q\!\prn{F_{\ms{-2}1}},
\label{Eq:ERrate}
\end{align}
where $F_{\ms{-2}1}$ takes on the value of the total angular
momentum of each ground state hyperfine level, $\ki$ is the
polarization-moment rank of the nucleus, and
\begin{align}\label{Eq:Pexpl}
    \mc{P}^{F}_{F_{\ms{-2}1}}\!\prn{\ki,\Gk} = & 3\sqrt{(2F_1+1)(2\ki+1)^2(2F+1)^3}~\times \\
        &~~~~~\ninej(\frac{1}{2},I,F)(\frac{1}{2},I,F)(1,\ki,\Gk)
            \ninej(\frac{1}{2},I,F_{\ms{-2}1})(\frac{1}{2},I,F_{\ms{-2}1})(1,\ki,\Gk).
            \nonumber
\end{align}
The terms in curly brackets in Eq.~\eqref{Eq:Pexpl} are nine-$J$
symbols \cite{Varshalovich}.

Spin-exchange collisions are electron-randomization collisions
among the alkali atoms. A key feature of spin-exchange collisions
is that because of angular momentum conservation, the overall
orientation of the alkali vapor must be preserved.  This leads to
an extra term
[$\Gd_{\Gk,1}\mc{P}^{F}_{F_{\ms{-2}1}}\!\prn{0,\Gk}$] on the
right-hand side of the equation \eqref{Eq:ERrate} describing the
relaxation of atomic polarization due to electron-randomizing
collisions \cite{OkunevichBook} [again, as in
Eq.~\eqref{Eq:ERrate}, $\kappa
> 0$]:
\begin{align}
    &\sbr{\frac{d}{dt}\rho^\Gk_q(F)}_\text{SE}
        = \label{Eq:SErate} \\
        &~~~~-\Gg_{se}
            \sum_{F_{\ms{-2}1}}
            \prn{\sum_{\ki}\mc{P}^{F}_{F_{\ms{-2}1}}\!\prn{\ki,\Gk}
            -\Gd_{\Gk,1}\mc{P}^{F}_{F_{\ms{-2}1}}\!\prn{0,\Gk}}
            \rho^{\Gk}_q\!\prn{F_{\ms{-2}1}}, \nonumber
\end{align}
where the spin-exchange rate is given by
$\gamma_{se}=n\sigma_{se}v\ts{rel}$, $n$ is the number density of
Cs atoms, $\sigma_{se}$ is the effective spin-exchange
cross-section, and $v\ts{rel}$ is the average relative velocity
between Cs atoms. Equation \eqref{Eq:SErate} is the linearized
version of the equations describing spin exchange, and is valid
when the orientation is sufficiently small. Experimental evidence
indicates that such an approximation is reasonable for our
experiment (see Sec.~\ref{SubSec:AssumptionsVerified}).

Finally, the uniform relaxation of atomic polarization with the rate $\gamma_u$ is
described by the equation
\begin{align}
    \sbr{\frac{d}{dt}\rho^\Gk_q(F)}_\text{U}
        =-\Gg_{u}\rho^{\Gk}_q\!\prn{F} ~~~~~~~~~~~ (\kappa > 0)~.
\end{align}

\begin{figure}
\includegraphics[width=3 in]{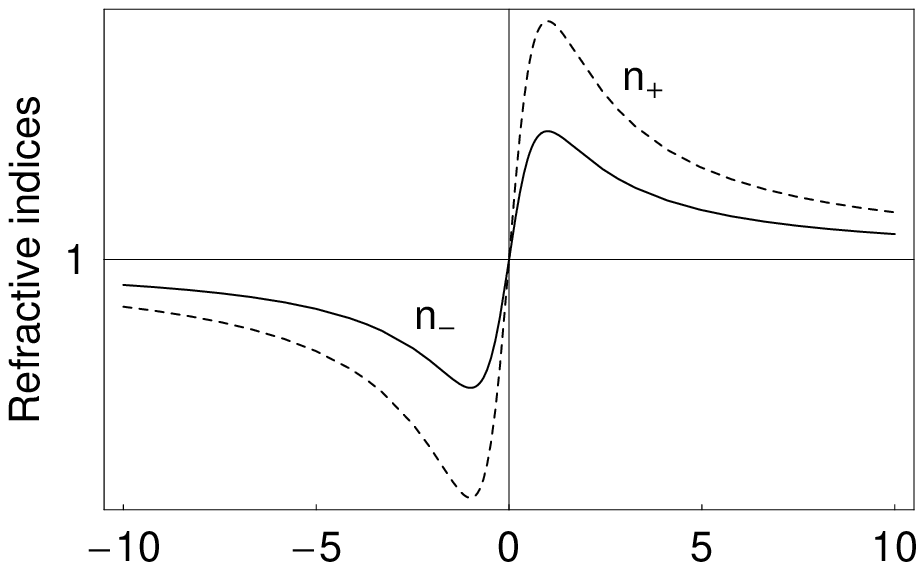}
\includegraphics[width=3 in]{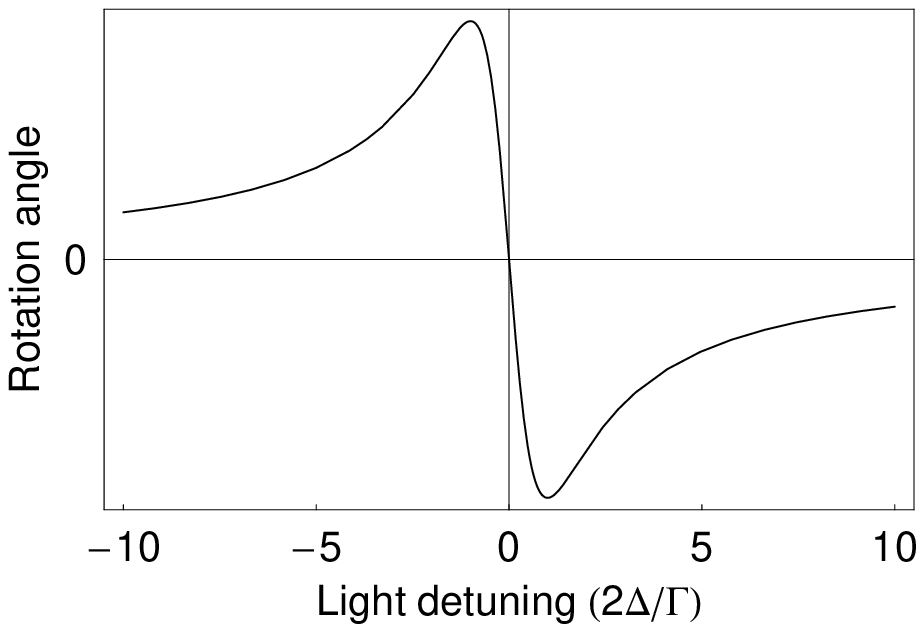}
\caption{Light frequency detuning $\Delta$ dependence of optical rotation (lower plot)
near an atomic resonance (occurring at $\Delta=0$).  The optical rotation is caused by a
difference in the amplitudes of the real parts of the complex indices of refraction
(upper plot) for left- and right-circularly polarized light ($n_+$ and $n_-$,
respectively).  The rotation angle is proportional to the difference $n_- - n_+$, since
it arises due to the difference in phase velocities between the circular components of
the linearly polarized light. For this plot, the magnetic field $B=0$, and a Lorentzian
model of line broadening is employed, where the width is $\Gamma$.}
\label{OpticalRotationMechanisms2}
\end{figure}

\begin{figure}
\includegraphics[width=3 in]{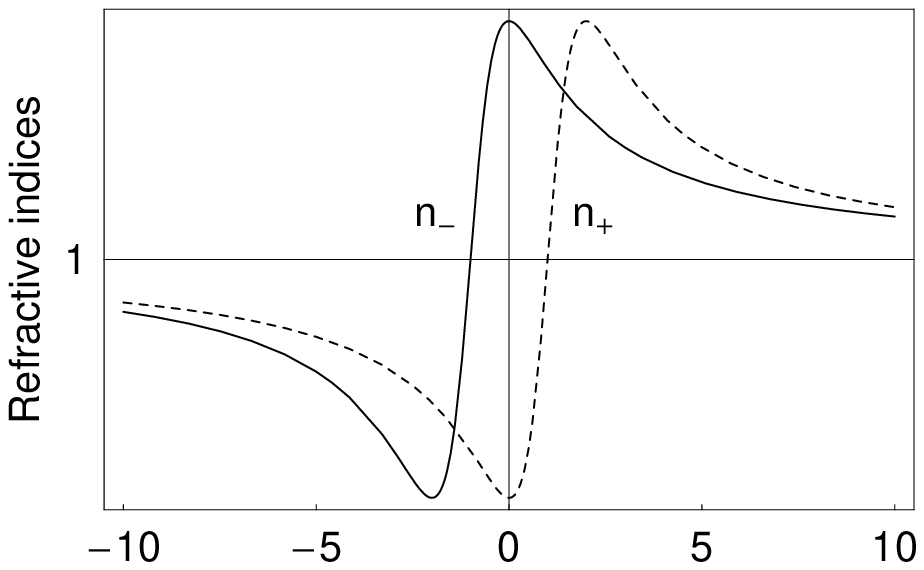}
\includegraphics[width=3 in]{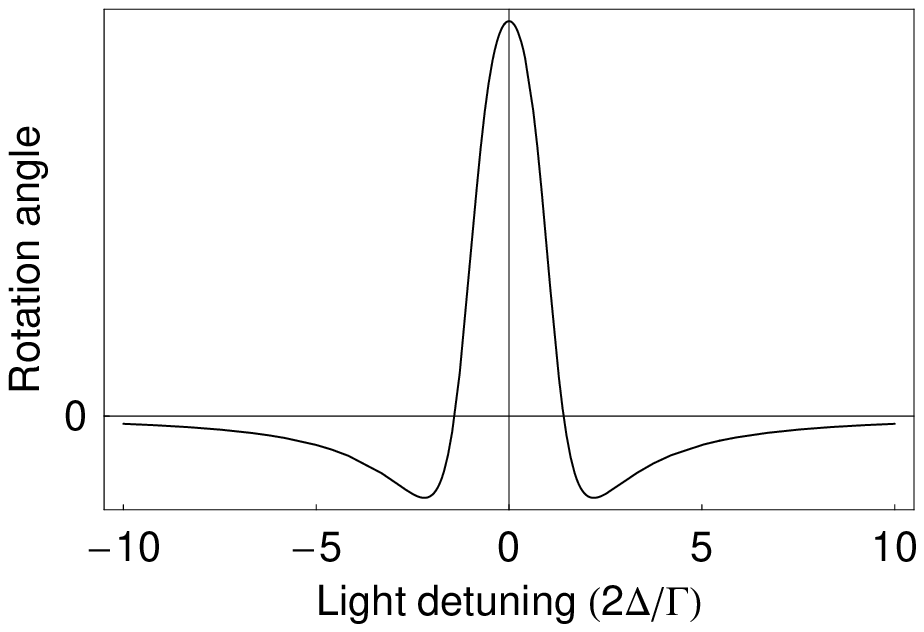}
\caption{Light frequency detuning $\Delta$ dependence of optical
rotation (lower plot) caused by splitting of the resonance
frequencies for left- and right-circularly polarized light due to
Zeeman shifts of sublevels in a magnetic field
$B=\Gamma/(2g\mu_0)$ ($\Gamma$ is the width of the resonance,
where we use a Lorentzian model of line broadening, $g$ is the
Land\'{e} factor, $\mu_0$ is the Bohr magneton). Upper plot shows
the dependence of the real parts of the complex indices of
refraction for left- and right-circularly polarized light in the
presence of a longitudinal magnetic field. The optical rotation is
proportional to the difference $n_- -
n_+$.}\label{OpticalRotationMechanisms1}
\end{figure}

Here we note that according to Eqs.~\eqref{Eq:ERrate} and \eqref{Eq:SErate}, to determine
the time dependence of a particular polarization moment for a given hyperfine level $F$,
one must solve two coupled linear differential equations (if both levels support the rank
$\kappa$ polarization moment being considered). Thus for alkali atoms, in order to find
the time dependence of a signal depending on the population, orientation, and alignment
in a given hyperfine level, it is necessary to employ three sets of two coupled linear
differential equations --- a total of six.  In general, a system of $N$ first-order
differential equations yields up to $N$ possible independent solutions \endnote{This can
be seen from the fact that one can solve such a system of equations using matrix methods
where the coefficients form a $N \times N$ matrix, which can have at most $N$ eigenvalues
and eigenvectors. These eigenvalues map directly to the solutions. However, it should be
noted that some of these solutions may be the same, leading to fewer than $N$ independent
time constants.}.

In the traditional Franzen's method of relaxation in the dark,
transmission of circularly polarized probe light is measured. The
absorption of circularly polarized light depends on all three
ground state polarization moments $\rho_0^{(0)}$, $\rho_0^{(1)}$,
and $\rho_0^{(2)}$ (intuitively this must be the case, since
circularly polarized light changes all three moments during the
optical pumping process). This leads to a time-dependent optical
signal involving many time constants.

However, if one measures optical rotation of narrow-band, linearly
polarized probe light (as in the scheme to measure relaxation in
the dark employed in the present experiment), at sufficiently low
magnetic fields, the signal is primarily sensitive only to the
orientation in the probed hyperfine level. Thus in the present
experiment, there appear only {\it{two}} time constants in the
optical signal.

Orientation produces optical rotation through a different physical mechanism than
population and alignment. Orientation in the probed hyperfine level creates a difference
in the amplitudes of the real parts of the complex indices of refraction for left- and
right-circularly polarized light (upper plot of Fig.~\ref{OpticalRotationMechanisms2}).
This causes optical rotation due to the difference in the phase velocities of the
circular components of the linearly polarized probe light, as illustrated in the lower
plot of Fig.~\ref{OpticalRotationMechanisms2}.  It is crucial to note that optical
rotation caused by longitudinal orientation appears even in the absence of a magnetic
field (and, in fact, is to first order independent of $B$).

On the other hand, in the absence of orientation, optical rotation
related to the $\kappa=0$ or $\kappa=2$ polarization moments
(population and alignment) appears only when a nonzero magnetic
field is present. The magnetic field splits the Zeeman sublevels,
producing a difference in the resonance frequencies of the real
parts of the complex indices of refraction for left- and
right-circularly polarized light
(Fig.~\ref{OpticalRotationMechanisms1}). This mechanism for
optical rotation of linearly-polarized light (whose frequency is
near-resonant with an atomic transition) is known as the
Macaluso-Corbino effect (see the review \cite{NMOEreview}). For
small magnetic fields, the amplitude of the optical rotation
signal due to the Macaluso-Corbino effect depends linearly on $B$.
This distinction, as well as the difference in optical rotation
spectra (see Figs.~\ref{OpticalRotationMechanisms2} and
\ref{OpticalRotationMechanisms1}), allows discrimination between
the signal caused by orientation and that caused by the
Macaluso-Corbino effect.

Here it should be noted that because cesium has nonzero nuclear
spin, mixing of different hyperfine components [with the same
$M_F$ value but different $F$] in the upper state of the
transition adds an important contribution to optical rotation due
to the Macaluso-Corbino effect
\cite{Khriplovich,TelegdiAndWeis,NMOEreview}. The rotation due to
this wavefunction-mixing effect also scales linearly with $B$ in
the low-field regime, but has a different dependence on light
detuning.  Nonetheless, because the antisymmetric-with-detuning
contributions of the different hyperfine components $F \rightarrow
F'$ are unresolved for $F'$ due to Doppler broadening, the overall
spectrum of time-dependent Macaluso-Corbino rotation due to
changes in population and alignment turns out to be quite similar
to that shown in Fig.~\ref{OpticalRotationMechanisms1}
\cite{TelegdiAndWeis}.

We can estimate the contribution of Macaluso-Corbino rotation to
the time-dependent optical rotation signal measured in our
experiment based on the amplitude of linear Faraday rotation that
is expected.  At the typical laser detuning and magnetic fields at
which we work ($\lesssim 15~{\rm{G}}$), taking into account the
efficiency of optical pumping, we expect a contribution of only a
few mrad to the time-dependent optical rotation amplitudes
\cite{TelegdiAndWeis}. Compared to the measured amplitudes of
time-dependent optical rotation due to changes in orientation
(20-80 mrad under usual conditions), this is a small correction
(on the order of the statistical noise in our measurements). This
is verified by measuring the dependence of the amplitudes as a
function of magnetic field (see
Sec.~\ref{SubSec:AssumptionsVerified}) and laser detuning (see
Sec.~\ref{SubSec:DetuningDependence}). Note, however, that the
contribution of the Macaluso-Corbino effect to time-dependent
rotation becomes important at higher magnetic fields.

Based on the above considerations, we assume that the
time-dependent optical rotation signal is due primarily to the
relaxation of atomic orientation along $z$. In our experiments,
the pump and probe light beams' frequencies are tuned to resonance
with the $F=4 \rightarrow F'$ hyperfine component of the D2
transition in Cs.  Therefore the optical rotation signal
$\varphi(t)$ in our experiments is proportional to
$\rho^{(1)}_0(F=4)$, and so according to the described theory for
the $F=4$ ground state hyperfine level of Cs ($I=7/2$),
$\varphi(t)$ is described by the following expression:
\begin{align}
\varphi(t) = \alpha_f e^{-\gamma_f t} + \alpha_s e^{-\gamma_s
t}+\varphi_0~, \label{Eq:FittingFunction}
\end{align}
where $\gamma_f$ and $\gamma_s$ are, respectively, the faster and slower rates of
relaxation given by
\begin{widetext}
\begin{align}
\gamma_{f,s} = \gamma_u + \frac{1}{64} \prn{33\gamma_{er} +
22\gamma_{se} \pm \sqrt{961\gamma_{er}^2 + 1324
\gamma_{er}\gamma_{se} + 484\gamma_{se}^2}}~,
\label{Eq:FastandSlowRates}
\end{align}
\end{widetext}
$\alpha_{f}$ and $\alpha_s$ are the respective amplitudes of the two exponentials, and
$\varphi_0$ is the dc rotation caused by the linear Faraday effect. Equation
\eqref{Eq:FittingFunction} is used to fit the obtained data to extract the relaxation
rates.

We can investigate several limits of the equations describing relaxation of atomic
orientation when various relaxation processes are not present.  According to
Eq.~\eqref{Eq:FastandSlowRates}, if $\gamma_{er}=0$ then the fast and slow rates are
given by:
\begin{align}
\gamma_f &= \gamma_u + \frac{11}{16} \gamma_{se}~, \\
\gamma_s &= \gamma_u~.
\label{Eq:NoER}
\end{align}
Since $\gamma_{se}=n\sigma_{se}v\ts{rel}$, this means that if $\gamma_{er}=0$, then one
expects the fast and slow rates to extrapolate to the same value for zero Cs density. If
$\gamma_{se}=0$, Eq.~\eqref{Eq:FastandSlowRates} yields for the fast and slow rates:
\begin{align}
\gamma_f &= \gamma_u + \gamma_{er}~, \label{Eq:RatesAtZeroDensity1} \\
\gamma_s &= \gamma_u + \frac{1}{32} \gamma_{er}~. \label{Eq:RatesAtZeroDensity2}
\end{align}
As we discuss in Sec.~\ref{Sec:RIDwithLIAD}, analysis of our experimental results yields
a nonzero value for $\gamma_{er}$, indicating that relaxation due to
electron-randomization collisions is prominent in the paraffin-coated cells studied. This
observation is consistent with other recent studies of relaxation in paraffin-coated
cells \cite{HexaDecapole,NISTpaper}.

\subsection{Experimental Setup}

\begin{figure}
\includegraphics[width=3.35 in]{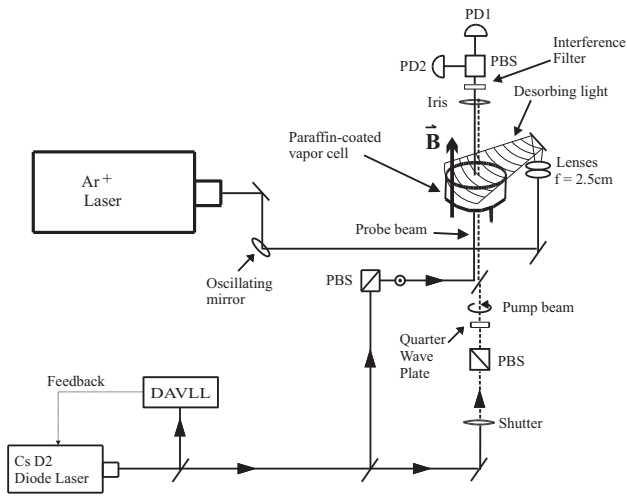}
\caption{Schematic diagram of the experimental setup. PBS -
polarizing beamsplitting cube, PD1(2) - photodiodes for light
detection in polarimeter, DAVLL - Dichroic Atomic Vapor Laser Lock
(described in text and shown in
Fig.~\ref{DAVLLsetup}).}\label{RIDsetup}
\end{figure}

The experimental apparatus for measuring spin relaxation is shown
in Fig.~\ref{RIDsetup}. The pump and probe beams, resonant with
the Cs D2 transition ($6s~^2S_{1/2}\rightarrow6p~^2P_{3/2}$), are
derived from the same 852-nm diode laser (Newport Model 2010
External Cavity Tunable Diode Laser).

\begin{figure}
\includegraphics[width=3.35 in]{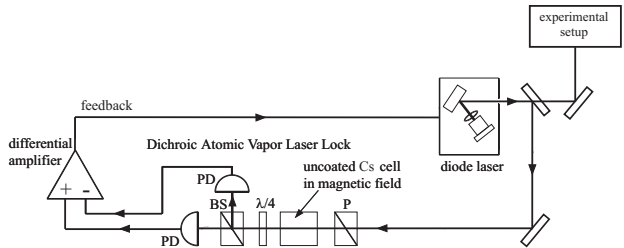}
\caption{Schematic of Dichroic Atomic Vapor Laser Locking (DAVLL)
system described in text. PD - photodiodes, BS - polarizing
beamsplitting cube, P - polarizer, $\lambda/4$ - quarter-wave
plate.  Detailed description can be found in Ref.~\cite{ourDAVLL}.
Light from the diode laser passes through a linear polarizer
before passing through an uncoated Cs cell at room temperature
($\approx 21^\circ{\rm C}$ -- corresponding to around one
absorption length for the center of the $F=4 \rightarrow F'$
hyperfine component of the D2 transition). The cell is immersed in
a magnetic field ($\sim 200~{\rm G}$, applied along the direction
of light propagation) that splits the Zeeman components of the
Doppler-broadened Cs absorption spectrum. An analyzer (that can be
continuously tuned from a circular to a linear analyzer) at the
output measures the resulting change in the light polarization
properties. The output of the analyzer functions as the error
signal for the electronic feedback system.}\label{DAVLLsetup}
\end{figure}

\begin{figure}
\includegraphics[width=3.35 in]{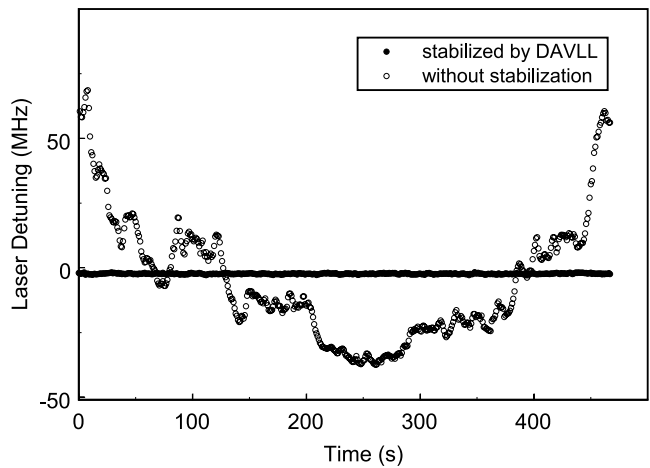}
\caption{Comparison of frequency stability of a Newport model 2010 external cavity diode
laser (central wavelength = 852 nm, tuned near the $F=4 \rightarrow F'$ hyperfine
component of the D2 line) with and without use of the DAVLL system \cite{ourDAVLL}. Laser
frequency is measured by saturated absorption spectroscopy using an uncoated Cs vapor
cell in an auxiliary setup.  The saturated absorption spectrum is power-broadened to
allow tracking of the laser detuning over the wide frequency range shown in the
plot.}\label{DAVLLstability}
\end{figure}

The frequency of the diode laser is controlled and monitored using
the Dichroic Atomic Vapor Laser Locking (DAVLL) system (see
Ref.~\cite{ourDAVLL} and references therein) illustrated in
Fig.~\ref{DAVLLsetup}.  In order to stabilize the laser frequency,
a small portion of the laser light is split off from the main beam
(light power $\sim 0.4~{\rm mW}$; beam diameter $\sim 3~{\rm mm}$)
and directed into the DAVLL.  The DAVLL system generates an
electronic feedback signal that is used to control the frequency
of the diode laser. Figure~\ref{DAVLLstability} shows the
improvement in the frequency stability of the laser when the DAVLL
system is employed. The DAVLL reduces the drift of the laser
frequency to $\lesssim 1~{\rm MHz}$ over the measurement time.

The main portion of the laser beam is split into a pump beam and a
probe beam (Fig.~\ref{RIDsetup}). The typical light power of the
pump beam is $4~{\rm mW}$ and the typical light power of the probe
beam is $3~{\rm \mu W}$, and their diameters are $\approx 3~{\rm
mm}$. The pump beam passes through a mechanical camera-iris type
shutter (that opens and closes at a rate of 0.25~Hz), and then
through a polarizing beam splitter followed by a quarter-wave
plate with fast axis at 45$^\circ$ to the axis of linear
polarization, which produces circular polarization. The normalized
Stokes parameter describing the degree of circular polarization
($S_2$, see, for example, Refs.~\cite{Huard}) is $ > 0.9$ for the
pump light. The light then passes through the paraffin-coated Cs
cell, but is blocked by an iris before it can hit the polarimeter
for the probe light (see below), thus avoiding saturation of the
photodiodes.

The probe is linearly polarized by a polarizing beam splitter and then directed along the
axis of the cell.  After passing through the cell, the probe light enters a balanced
polarimeter which measures its optical rotation. The polarimeter is fitted with an
interference filter centered at 850 nm (with a bandwidth of 12 nm FWHM) to eliminate
detection of scattered light from the Ar$^+$ laser (used for experiments with LIAD, see
below). We record both the sum ($P_1+P_2$) and the difference ($P_1-P_2$) of the
photodiode signals from the polarimeter. The rotation angle is found according to:
\begin{align}
\varphi = \frac{P_1-P_2}{2(P_1+P_2)}~. \label{Eq:RotationAngle}
\end{align}
The sum signal is a measure of light transmitted through the cell.  By scanning the
frequency of the laser light, the sum signal gives the absorption spectrum which can be
fit to a sum of Voigt profiles -- allowing one to calculate the vapor density of Cs in
the cell.

The three paraffin-coated vapor cells studied in this work are
cylindrical glass cells with dimensions listed in
Table~\ref{Table:CellDimensions}. Each cell has a single stem
containing a droplet of Cs metal.  The stems have circular
openings of diameter $\sim 0.3~{\rm mm}$, although the diameters
of the openings vary from cell to cell by up to a factor of two.
The size of the hole is chosen to be small enough that relaxation
due to the ``reservoir effect'' \cite{Bouchiat,BouchiatPhD} is
small compared to other sources of relaxation. The cells are
evacuated to a residual pressure of $\approx 10^{-5}~{\rm Torr}$
during manufacture and are nominally free of any buffer gas.
Detailed information on the manufacture and properties of similar
cells can be found in Ref.~\cite{ourLIAD}.

\begin{table}
\caption{Dimensions of the cylindrical paraffin-coated Cs vapor
cells used in this work.  The fourth column lists the mean free
path of an atom between wall collisions for the particular
geometry of the cell, determined by a Monte-Carlo simulation that
assumes a cosine distribution of atoms reflected from the paraffin
surface.}
\medskip \begin{tabular}{cccc} \hline \hline
Cell~ & ~diameter (cm)~ & ~length (cm)~ & ~mean free path (cm)\\
\hline A & 6 & 3 & 3.4 \\
B & 6 & 3 & 3.4 \\
C & 2 & 2 & 1.5 \\
\hline \hline
\end{tabular}
\label{Table:CellDimensions}
\end{table}

A magnetic field directed along the axis of light propagation of
up to $\sim 15~{\rm G}$ is applied to the cell with a pair of
Helmholtz coils.  For an applied field of $\sim$1~G, the variation
of the longitudinal and transverse components of the field over
the volume of the cell were less than 3\% of the magnitude of the
leading field.  The field variation was measured with a flux-gate
magnetometer mounted on a translation stage movable in three
dimensions, and the measurement was carried out at $\sim$1~G to
accommodate the magnetic field range of the flux-gate
magnetometer.  We expect inhomogeneities of the magnetic field to
either scale proportionally to the applied field (for example, if
they are due to coil geometry) or be reduced at higher magnetic
fields (for example, if they are due to a stray inhomogeneous
magnetic field generated by some object in the laboratory).  Thus
we assume that the field homogeneity in our experiments was better
than 3\%.

The Ar$^+$ laser (lasing on the $514~{\rm nm}$ line) is employed
in the measurements of relaxation in the dark during LIAD. The
laser beam is reflected from a vibrating mirror which serves to
average most of the interference pattern in the laser light
(speckle), and expanded using lenses in order to illuminate the
entire cell. The intensity of light incident on the cell ranged
from $1.1~{\rm mW/cm^2}$ to $110~{\rm mW/cm^2}$.

\subsection{Results and Discussion}\label{Sec:RIDOVORresults}

\begin{figure}
\includegraphics[width=3.35 in]{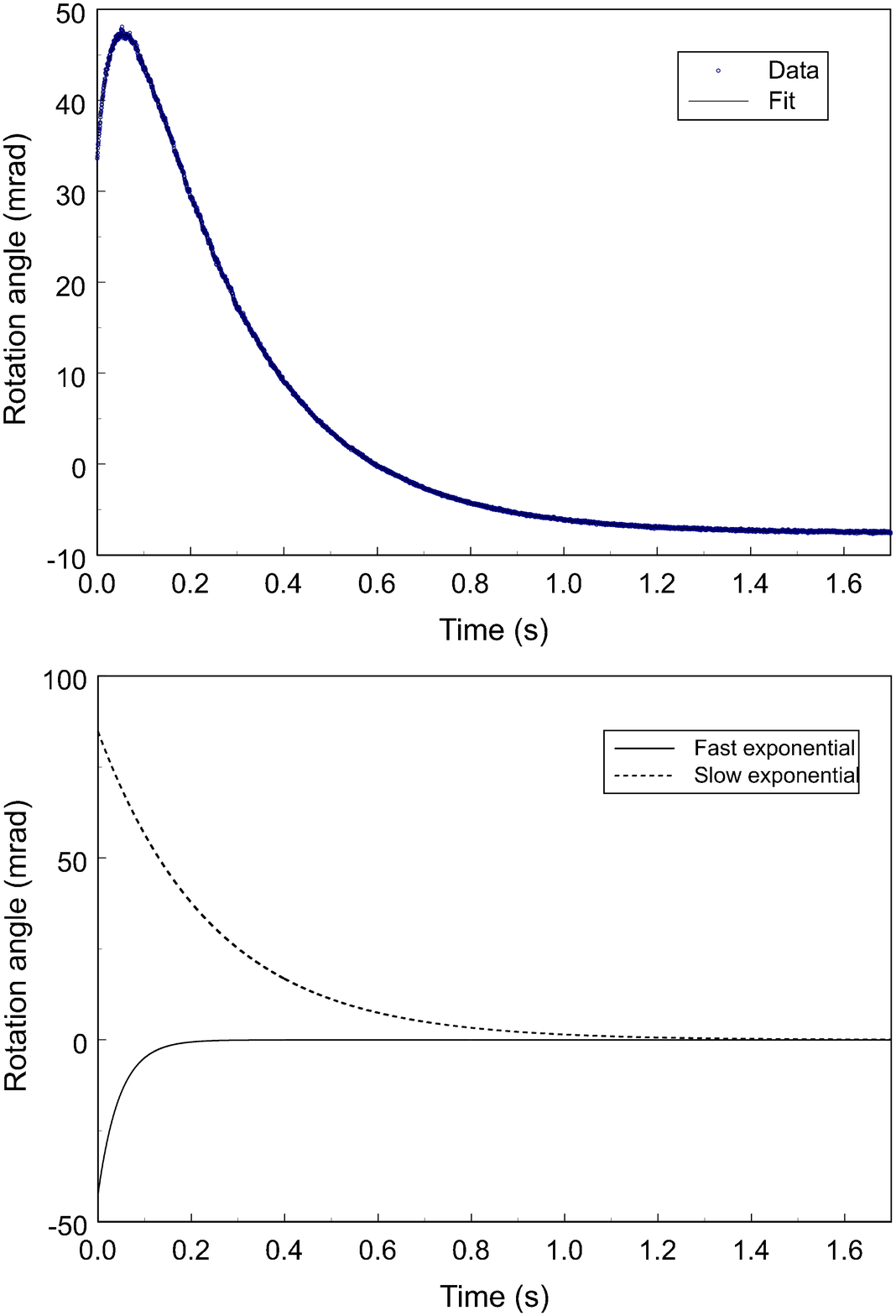}
\caption{Upper plot shows typical data for the time-dependent
component of optical rotation of the linearly polarized probe
light after the circularly polarized pump light is blocked. There
is clear evidence of two oppositely signed contributions to
optical rotation that relax at different rates, shown in the lower
plot. Cell temperature 21$^\circ$C, Cs density = $1.7\times
10^{10}~{\rm atoms/cm^3}$, pump light power = 4 mW, probe light
power = 3 $\mu$W, and $|\vec{B}| \approx 2~{\rm G}$. Data taken
with cell A (Table~\ref{Table:CellDimensions}). Diode laser is
tuned about 400~MHz to the low frequency side from the center of
the $F=4 \rightarrow F'$ hyperfine component of the Cs D2
absorption line, where the amplitudes of the two contributions are
large (see Fig.~\ref{AmpRateSpectrum}).  Lower plot shows the two
oppositely signed contributions to the optical rotation signal and
their exponential decay (extracted from the fit to the data shown
in the upper plot).}\label{SampleData}
\end{figure}

\subsubsection{Time-dependent optical rotation}
\label{SubSec:TimeDepOptRot}

Typical data for time-dependent optical rotation is shown in Fig.~\ref{SampleData}. This
is a measurement of the rotation angle according to Eq.~\eqref{Eq:RotationAngle} during
the time when the circularly polarized pump light is blocked by the shutter. The data
indicates two contributions to optical rotation with opposite signs that relax at
different rates, allowing us to determine $\gamma_f$ and $\gamma_s$ according to
Eq.~\eqref{Eq:FittingFunction}.

It may at first glance seem strange that relaxation of atomic polarization can cause the
magnitude of optical rotation (and by inference, the amount of orientation in the probed
level) to first increase, and then decrease in time (Fig.~\ref{SampleData}). The optical
pumping process creates orientation in both the $F=4$ and $F=3$ hyperfine levels, but the
probe beam only measures the orientation in the $F=4$ level. Uniform relaxation, which is
the dominant contribution to the slow relaxation rate $\gamma_s$
[Eq.~\eqref{Eq:FastandSlowRates}], reduces the degree of orientation in both hyperfine
levels.  However, electron randomization and spin-exchange collisions (which dominate
$\gamma_f$) transfer orientation from one level to another, and thus can increase the
orientation in the probed level.

\subsubsection{Verification of assumptions in model describing optical rotation signal: light-power and magnetic-field
dependences}\label{SubSec:AssumptionsVerified}

In order to apply the analysis outlined in Sec.~\ref{SubSec:PrincipleOfRIDOVOR}, it is
crucial to verify that the experiments were performed under conditions consistent with
the assumptions of the model.

\begin{figure}
\includegraphics[width=3.35 in]{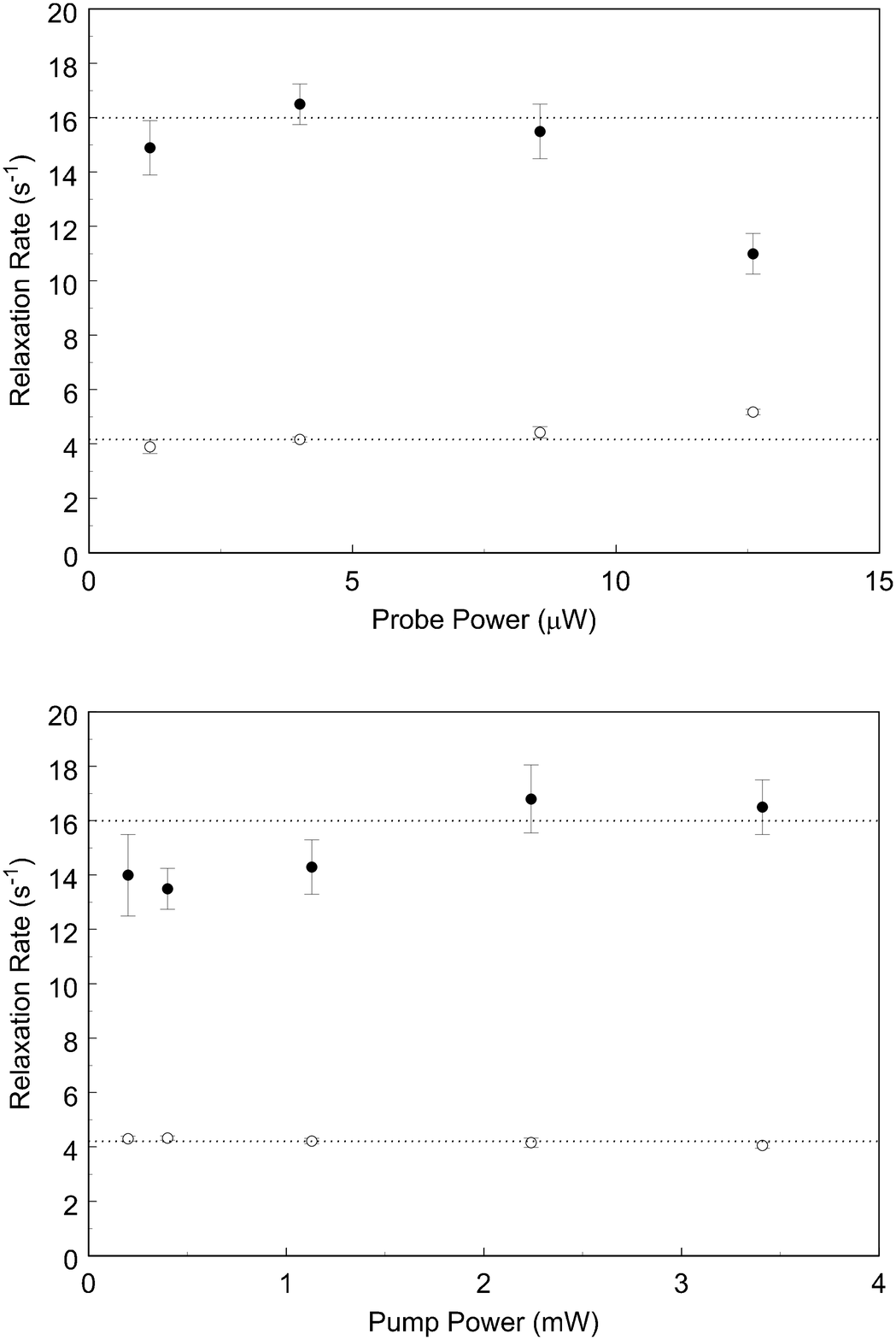}
\caption{Relaxation rates $\gamma_s$ (open circles) and $\gamma_f$
(filled circles) as a function of probe light power (upper plot)
and pump light power (lower plot). Dashed lines represent the
average values of the relaxation rates for the light powers where
all other data is acquired (pump power $\approx 4~{\rm mW}$, probe
power $\approx 3~{\rm \mu W}$). Data taken with cell B
(Table~\ref{Table:CellDimensions}).  Diode laser is tuned about
400~MHz to the low frequency side from the center of the $F=4
\rightarrow F'$ hyperfine component of the Cs D2 absorption line
(see Fig.~\ref{AmpRateSpectrum}), cell temperature $\approx
21^\circ$C, Cs density = $8\times 10^{9}~{\rm atoms/cm^3}$,
$|\vec{B}| \approx 10~{\rm G}$. Data points reflect the average of
data for several measurements. Light power was changed by
inserting various neutral density filters into the beam paths.
Above probe light powers of $\sim 8~{\rm \mu W}$, evidence of
optical pumping is seen as the relaxation rates are clearly
affected. All other measurements reported in this work are taken
with probe-light powers around $3~{\rm \mu W}$.}\label{PowDep}
\end{figure}

First, for the purposes of our measurements, it is important that
the probe beam does not cause any optical pumping and thus does
not affect the spin-relaxation rates we seek to measure. As the
data in the upper plot of Fig.~\ref{PowDep} demonstrates, the
light power of the probe beam has no significant effect on the
relaxation rates as long as the power is kept below $\sim 8~\mu
W$.  In all other data, the probe beam's power is kept well below
this level so that it measures the orientation of the Cs gas
without disturbing the atomic polarization during the part of the
experiment that is meant to be ``dark."

Second, the linearization of the spin-exchange equations [Eq. \eqref{Eq:SErate}] hinges
on the assumption of small orientation, placing a limit on the power of the pump beam.
At the same time, the pump beam must be sufficiently intense to produce a measurable
signal. The lower plot of Fig.~\ref{PowDep} shows that the relaxation rates are
relatively independent of light power (within 10\% of the mean value) over the range of
powers measured, indicating that Eq.~\eqref{Eq:SErate} is adequate for the description of
our data.  (The slight increase ($\sim 1-2~{\rm s^{-1}}$) in relaxation at high pump
light powers may be a hint of increased spin relaxation due to violation of the small
orientation condition.)

Third, it is essential that the field is large enough that the
optical rotation signal is not affected by precession of atomic
polarization in stray fields and that spin relaxation due to
magnetic field gradients can be neglected (as discussed below). On
the other hand, as was discussed in detail in
Sec.~\ref{SubSec:PrincipleOfRIDOVOR}, the longitudinal magnetic
field applied by the Helmholtz coils cannot be so large that the
optical-rotation signal becomes sensitive to atomic polarization
moments other than orientation (the $\kappa =1$ moment).
Time-dependent optical rotation due to other polarization moments
comes about because of the Macaluso-Corbino effect (see the review
\cite{NMOEreview}).  For example, since the laser light in our
experiment is resonant with one particular ground-state hyperfine
level, if optical pumping changes the population of that hyperfine
level, then Macaluso-Corbino rotation changes amplitude in time
due to the relaxation of the population (the $\kappa=0$ moment)
difference ``in the dark.''

\begin{figure}
\includegraphics[width=3.35 in]{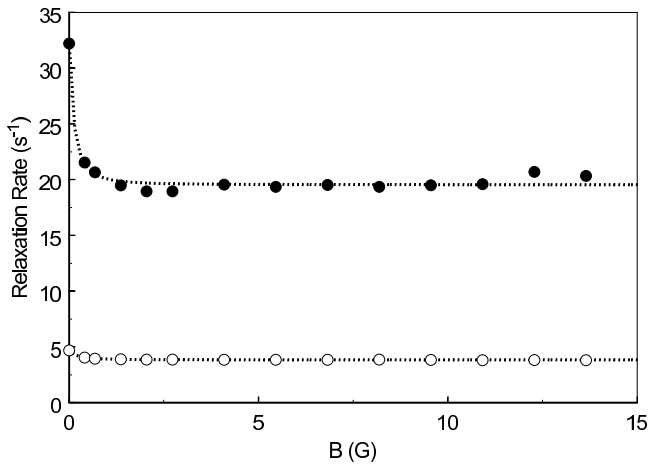}
\caption{Magnetic-field dependence of the relaxation rates
$\gamma_s$ (open circles) and $\gamma_f$ (filled circles). Data
taken with cell B (Table~\ref{Table:CellDimensions}). Laser
frequency detuned $\approx 400~{\rm MHz}$ to the high-frequency
wing of of the $F=4 \rightarrow F'$ hyperfine component of the Cs
D2 absorption line (where the amplitudes of the time-dependent
signals are large, see Figs.~\ref{AmpRateSpectrum} and
\ref{AmpRateSpectrum}). Pump light power = 4 mW, probe light power
= 3 $\mu$W, Cs density = $8 \times 10^9~{\rm
atoms/cm^3}$.}\label{BfieldDependence}
\end{figure}

For most data a magnetic field of $\approx 7$ G was applied along
the direction of light propagation. This field served as the
leading magnetic field $B$ along which the atomic orientation was
directed.  The stray transverse field in the laboratory (due
primarily to the Earth's magnetic field and magnetic properties of
the optical table) at the position of the vapor cell was measured
to be $\approx 0.3~{\rm G}$ and directed vertically. The data
presented in Fig.~\ref{BfieldDependence} show that relaxation
rates level off at fields above $\sim 1.5$~G, demonstrating that
when the applied longitudinal field sufficiently exceeds the stray
transverse magnetic field, the rates are field-independent (over
this range). For smaller values of the applied magnetic field, the
stray field significantly tilts the total magnetic field vector
away from the light propagation direction. In this case precession
of the oriented atoms tends to average out the net atomic
polarization. This is the reason for the apparent increase in
$\gamma_s$ and $\gamma_f$ for $B \lesssim 1.5$~G.

\begin{figure}
\includegraphics[width=3.35 in]{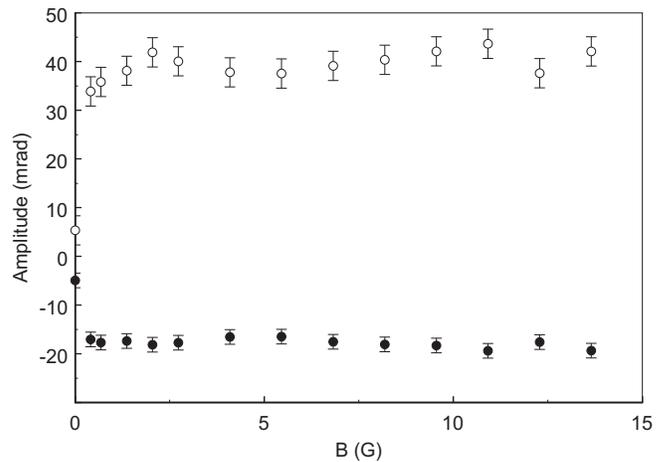}
\caption{Magnetic-field dependence of the amplitudes of
time-dependent optical rotation $\alpha_s$ and $\alpha_f$ [see
Eq.~\eqref{Eq:FittingFunction}], same conditions as in
Fig.~\ref{BfieldDependence}. Open circles correspond to
$\alpha_s$, filled circles correspond to
$\alpha_f$.}\label{BfieldDependenceAmp}
\end{figure}

Figure \ref{BfieldDependenceAmp} shows the magnetic field dependence of the amplitudes of
time-dependent optical rotation, $\alpha_s$ and $\alpha_f$ [see
Eq.~\eqref{Eq:FittingFunction}].  We observe a dramatic change in the amplitudes for
magnetic fields below $\sim 1.5$~G, due to the influence of the stray laboratory field
discussed above. Above this value, there appears to be no linear dependence of the
amplitudes on magnetic field over the studied range.  This is a key test verifying that
the optical rotation signal is dominated by the $\kappa=1$ moment and relatively
insensitive to the Macaluso-Corbino effect at the light detuning and magnetic field
conditions at which we work.

The data shown in Figs.~\ref{BfieldDependence} and
\ref{BfieldDependenceAmp} also reveal how the relaxation rates and
optical rotation signal amplitudes depend on the alignment of the
leading magnetic field with the light propagation direction.  For
our experimental geometry (Fig.~\ref{RIDsetup}), we have verified
that the light propagation directions of the pump and probe beams
are collinear with the applied magnetic field to within 2$^\circ$.
As the amplitude of the applied magnetic field is increased from
zero, the field goes from nearly perpendicular to the light
propagation direction to nearly parallel.  From the data plotted
in Figs.~\ref{BfieldDependence} and \ref{BfieldDependenceAmp}, we
conclude that uncertainty in the alignment of the light beams and
the magnetic field is a negligible contribution to the uncertainty
in the determination of the relaxation rates and optical rotation
amplitudes.

Under the conditions of our experiment, relaxation due to
magnetic-field gradients is negligible.  This can be seen as
follows.  The presence of gradients can be modelled by assuming
there is a small transverse field $\Delta\vec{B}$ in one half of
the cell (the magnitude of the transverse gradient field is much
smaller than the leading field, $|\Delta\vec{B}| \ll |\vec{B}|$).
Between collisions with the cell wall, the Cs atoms' orientation
adiabatically follows the direction of the total magnetic field
($\vec{B}_{tot} = \vec{B} + \Delta\vec{B}$).  This is assured by
the fact that $\Omega_L\gg\upsilon/R$ where $\Omega_L=\gamma B$ is
the Larmor frequency ($\gamma$ is the gyromagnetic ratio), $R$ is
the characteristic dimension of the vapor cell and $\upsilon$ is
the atoms' thermal velocity. Collisions with the wall break this
adiabatic condition, as discussed in detail in
Refs.~\cite{HapperReview,OurBook}, leading to spin relaxation
described by \cite{OurBook}:
\begin{align}
    \frac{1}{T_1} \sim \prn{\frac{\Delta B}{\gamma B^2}\frac{\upsilon}{R}}^2
    \frac{\upsilon}{R}~,
\label{Eq:GradientRelax}
\end{align}
where $T_1$ is the longitudinal spin-relaxation time. From
Eq.~\eqref{Eq:GradientRelax} we find that the relaxation rate due
to gradients is completely negligible under the conditions of this
work ($T_1 \sim 10^6~{\rm s}$ at $B \sim 10~{\rm G}$).  This
conclusion is further substantiated by the data presented in
Fig.~\ref{BfieldDependence}, which demonstrates that there is no
discernible field dependence of the relaxation rates over the
range of fields at which we work.

\subsubsection{Laser detuning dependence}\label{SubSec:DetuningDependence}

\begin{figure}
\includegraphics[width=3.35 in]{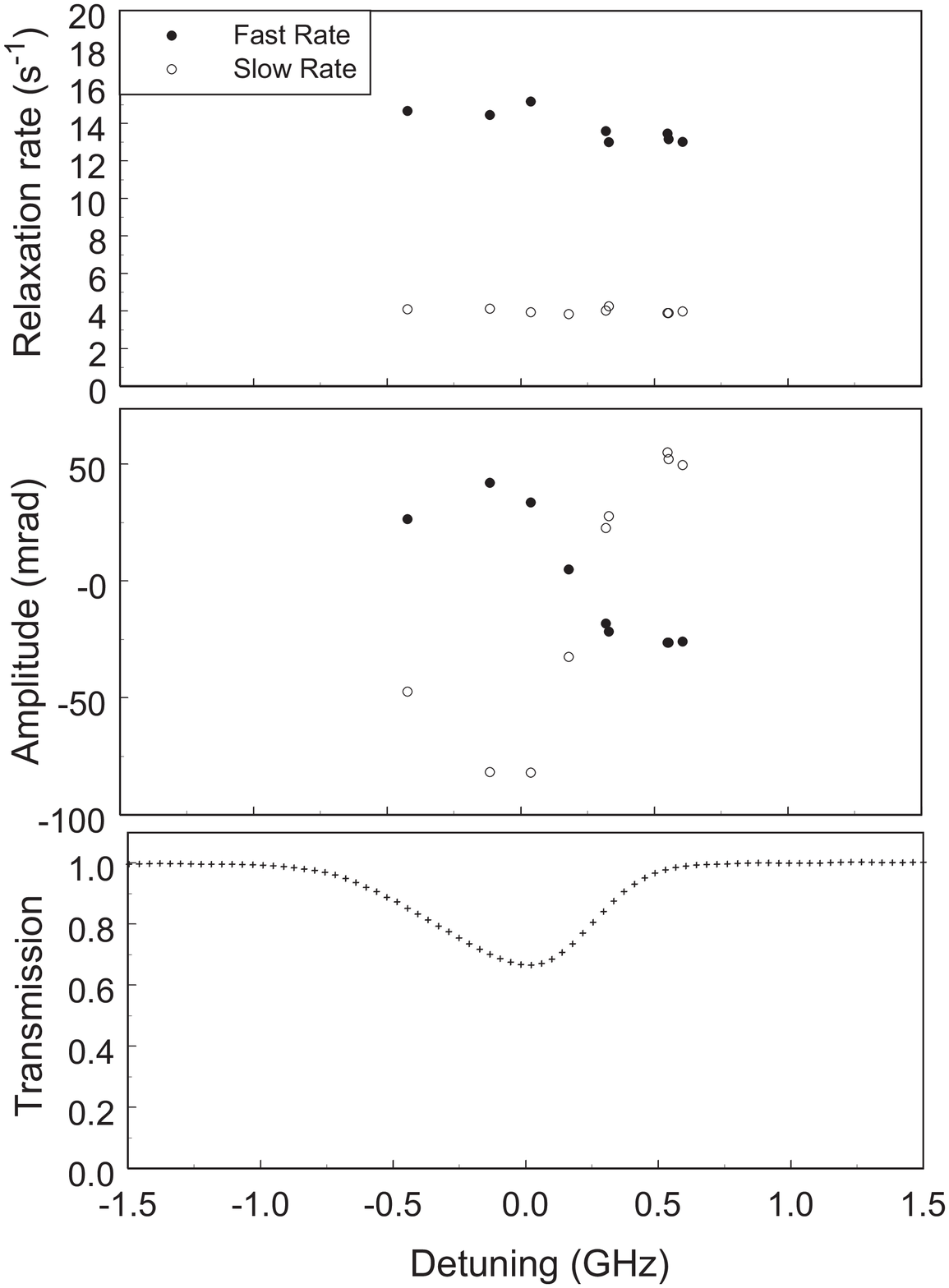}
\caption{Dependence of the relaxation rates (upper plot) and
amplitudes of optical rotation (middle plot) on detuning from
resonance. Zero detuning corresponds to the center of the $F=4
\rightarrow F'$ hyperfine component of the Cs D2 absorption line.
Pump light power = 4 mW, probe light power = 3 $\mu$W, Cs density
= $8 \times 10^9~{\rm atoms/cm^3}$, $|\vec{B}|=10~{\rm G}$. Open
circles correspond to the slower rate $\gamma_s$ and the
corresponding rotation amplitude $\alpha_s$, filled circles
correspond to the faster rate $\gamma_f$ and the corresponding
rotation amplitude $\alpha_f$ [see
Eq.~\eqref{Eq:FittingFunction}]. Data taken with cell B
(Table~\ref{Table:CellDimensions}). Lower plot shows the
transmission spectrum for the low-power probe light in the absence
of pump light and magnetic field.}\label{AmpRateSpectrum}
\end{figure}

The upper plot in Fig.~\ref{AmpRateSpectrum} illustrates the
laser-detuning dependence of the two relaxation rates ($\gamma_s$
and $\gamma_f$) near the Doppler-broadened $F=4 \rightarrow F'$
hyperfine component of the Cs D2 line. The relaxation rates are
relatively independent of detuning (deviations from the average
values are less than $10\%$). The small detuning dependence of the
relaxation rates seen in the data shown in
Fig.~\ref{AmpRateSpectrum} may be due to a slight violation of the
small orientation condition for linearized spin exchange assumed
in the derivation of Eq.~\eqref{Eq:SErate}. Another possible cause
of the slight detuning dependence is a small contribution to the
rates from the Macaluso-Corbino effect.

The middle plot in Fig.~\ref{AmpRateSpectrum} shows the dependence
of the amplitudes of the optical rotation signals on laser
detuning. The dispersive character of the spectrum of the
amplitudes $\alpha_s$ and $\alpha_f$ seen in
Fig.~\ref{AmpRateSpectrum} is roughly what would be expected from
optical rotation produced by an oriented sample of atoms --- the
dispersive function shown in Fig.~\ref{OpticalRotationMechanisms2}
must be convolved with a Gaussian function to account for Doppler
broadening and multiplied by a Voigt lineshape function to account
for optical pumping (as well as taking into account the unresolved
hyperfine structure). Optical rotation related to the
Macaluso-Corbino effect would have a mostly symmetric spectrum
derived from that shown in Fig.~\ref{OpticalRotationMechanisms1}.

One may notice that the centers (zero-crossings) of the
dispersively shaped spectra of the optical rotation amplitudes
$\alpha_s$ and $\alpha_f$ are shifted by around 250 MHz from the
center of the low light power transmission spectrum in
Fig.~\ref{AmpRateSpectrum}. This may be explained by recalling
that on the high frequency side of the Doppler-broadened resonance
the pump and probe light are resonant with the $F=4 \rightarrow
F'=5$ cycling transition, which may enhance the contribution of
this hyperfine component to the optical rotation signal.
Furthermore, it is important and convenient that the two
amplitudes $\alpha_s$ and $\alpha_f$ generally are of opposite
sign.  A possible reason for this difference in sign was offered
in Sec.~\ref{SubSec:TimeDepOptRot} -- namely that
electron-randomization collisions and spin-exchange collisions,
which dominate the fast rate $\gamma_f$, can transfer orientation
from the unprobed $F=3$ ground state hyperfine level to the probed
$F=4$ level, while uniform relaxation, which is the dominant
contribution to $\gamma_s$, decreases orientation in both levels.
Therefore, in principle, $\gamma_f$ can be related to an increase
in the orientation in the $F=4$ level and $\gamma_s$ can be
associated with a decrease in the orientation in the $F=4$ level,
which in turn would cause the amplitudes $\alpha_s$ and $\alpha_f$
to have opposite signs.

A complete density-matrix calculation describing the optical
pumping, evolution of atomic polarization, and optical probing of
alkali atoms contained in paraffin-coated cells is in progress.

\subsubsection{Cell-temperature dependence}
\label{SubSec:CellTempDep}

\begin{figure}
\includegraphics[width=3.35 in]{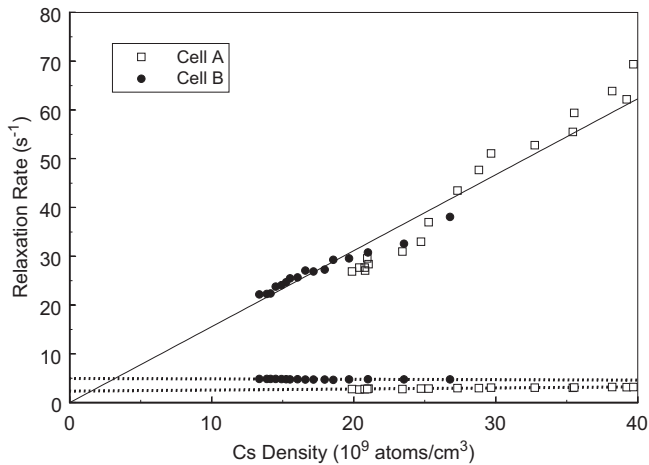}
\caption{Relaxation rates ($\gamma_f$ and $\gamma_s$) as the
temperature of cells A and B (Table~\ref{Table:CellDimensions}) is
changed, plotted with respect to the Cs density. Cell A's
temperature was varied from $\approx 24.3^\circ{\rm C}$ to
$\approx 30^\circ{\rm C}$, while cell B's temperature was varied
from $\approx 22.4^\circ{\rm C}$ to $\approx 27.6^\circ{\rm C}$. A
single linear fit describes the fast rates for both cells (solid
line), separate linear fits are carried out for the slow rates
(dashed lines). Laser frequency is detuned $\approx 400~{\rm MHz}$
to the high-frequency wing of of the $F=4 \rightarrow F'$
hyperfine component of the Cs D2 absorption line (where the
amplitudes of the time-dependent signals are large, see
Figs.~\ref{AmpRateSpectrum} and \ref{AmpRateSpectrum}), pump light
power = 4 mW, probe light power = 3 $\mu$W, $|\vec{B}|=10~{\rm
G}$.  It is apparent that there is some systematic deviation of
the fast relaxation rate data from the linear trend (most of the
data for $\gamma_f$ at Cs densities between $20-30 \times
10^{9}~{\rm{cm^{-3}}}$ fall below the linear fit, while most of
the data for Cs densities $> 30 \times 10^{9}~{\rm{cm^{-3}}}$
falls above the linear fit).  The spread of these systematic
deviations is on the order of the run-to-run reproducibility of
the data, so at present we do not ascribe it any particular
significance.}\label{TempDep}
\end{figure}

Figure~\ref{TempDep} illustrates the atomic polarization
relaxation rates in cells A and B
(Table~\ref{Table:CellDimensions}) as a function of Cs density
--- the density was altered by varying the ambient air temperature
in an insulated foam box containing the vapor cell. The
temperature dependence of the fast rate of relaxation $\gamma_f$
is roughly similar for both cells. Since, according to the model
described in Sec.~\ref{SubSec:PrincipleOfRIDOVOR}, $\gamma_f$ is
dominated by electron-randomization and spin-exchange collisions
(which depend only on the Cs density), we see that our data is
consistent with the observation that paraffin-coated cells of
similar construction and size have similar values for
$\gamma_{er}$ \cite{NISTpaper}. The slow rate, consistent with the
hypothesis that $\gamma_s$ is dominated by the $\gamma_u$ (caused
by the reservoir effect), does appear to depend on the specific
cell (presumably because the size of the stem opening varies
between cells) but not on the temperature.

A linear fit of the data describing the fast rate of relaxation
with respect to Cs density yields a slope of $1.56(2) \times
10^{-9}~{\rm cm^3/s}$.  The spin exchange rate $\gamma_{se}$ is
given by $\gamma_{se} = n\sigma_{se}v\ts{rel}$, where $n$ is the
density of Cs atoms, $\sigma_{se} \approx 2\times10^{-14}~{\rm
cm^2}$ \cite{Bouchiat} is the spin exchange cross section, and
$v\ts{rel} \approx 3\times 10^4~{\rm cm/s}$ is the average
relative velocity between the Cs atoms. Using this relation for
$\gamma_{se}$ in Eq.~\eqref{Eq:FastandSlowRates}, the contribution
to the slope of the fast rate in Fig.~\ref{TempDep} from spin
exchange is $\approx (11/16)\gamma_{se} \approx 0.4 \times
10^{-9}~{\rm cm^3/s}$, i.e. about a factor of $4$ smaller than the
observed slope. This means that the fast relaxation rate is not
dominated by spin-exchange collisions, but instead by electron
randomization collisions with the wall or, perhaps, some gaseous
impurities.

The relaxation due to electron randomization collisions evidently depends on the cell
temperature or the cesium vapor density, and in the latter scenario has a nearly linear
dependence on Cs density.  The origin of this relaxation is not presently understood.

One basic question is whether or not such relaxation has been
observed in other antirelaxation coated cells, and what
information can be drawn from these previous studies.

In Ref.~\cite{NISTpaper}, measurements of the widths and frequency
shifts for microwave transitions in paraffin-coated rubidium cells
are compared to data on Zeeman relaxation obtained from nonlinear
magneto-optical rotation measurements.  In that work, there
emerges compelling evidence that electron-randomization collisions
on the wall dominate spin-relaxation, in agreement with our
findings at room temperature and above.

In Ref.~\cite{Vanier}, the authors measured relaxation rates
associated with the 3.03 GHz $^{85}$Rb 0-0 hyperfine transition in
a Paraflint coated cell as a function of cell temperature. They
took measurements for different stem temperatures and extrapolated
to zero Rb density to isolate effects dependent on the coating
temperature from effects dependent on the Rb density.  In this
case electron randomization collisions were not found to dominate
relaxation. This may be a hint that it is in fact some
modification of the coating surface by the alkali atoms that is
responsible for the relaxation due to electron randomization
collisions in our experiment.  In fact, experiments carried out by
the same group discussed in Ref.~\cite{Boulanger} seem to hint
that in a similar situation when the data is not extrapolated to
zero alkali density, electron randomization collisions do in fact
dominate relaxation.

In Ref.~\cite{AleksandrovK}, the spin-relaxation effects in a paraffin-coated cell
containing potassium were measured.  Based on measurements of the broadening of a
potassium magnetic resonance line as a function of cell temperature and potassium vapor
density, it was determined that spin-exchange collisions between the potassium atoms were
the dominant source of relaxation.  This may indicate that the relaxation due to electron
randomizing wall collisions observed in Ref.~\cite{NISTpaper} and the present work are
somehow specific to rubidium and cesium.

Thus it is apparent that according to available literature, the
nature of this relaxation is unclear at the present time, and
further experimentation is warranted.

\section{Relaxation of atomic polarization in the presence of light-induced atomic desorption}
\label{Sec:RIDwithLIAD}

\setcounter{subsubsection}{0}

Alkali atoms in paraffin-coated vapor cells are absorbed into the
cell coating over time. When these cells are then exposed to light
of sufficiently short wavelength, alkali atoms are desorbed from
the paraffin coating into the volume of the cell \cite{ourLIAD}.
This phenomenon, known as light-induced atomic desorption (LIAD),
has been observed using a wide range of surfaces: sapphire
\cite{AlexandrovLIAD,BonchB1,BonchB2}, silane-coated glass (in
particular poly-dimethylsiloxane)
\cite{PDMS_gozzini,PDMS_meucci,PDMS_mariotti,PDMS_xu,PDMS_atutov},
superfluid $^4$He films \cite{Yabuzaki,YabuzakiNew}, quartz
\cite{QuartzLIAD}, and porous silica \cite{PorousSilicaLIAD}. LIAD
is useful as a method for the rapid control of atomic density, and
is of particular interest in the development of miniaturized
atomic clocks and magnetometers \cite{NISTpaper}.  A primary
question is whether or not LIAD affects the relaxation properties
of the wall coating. Additionally, can LIAD be used to understand
the relaxation processes?

\begin{figure}
\includegraphics[width=3.35 in]{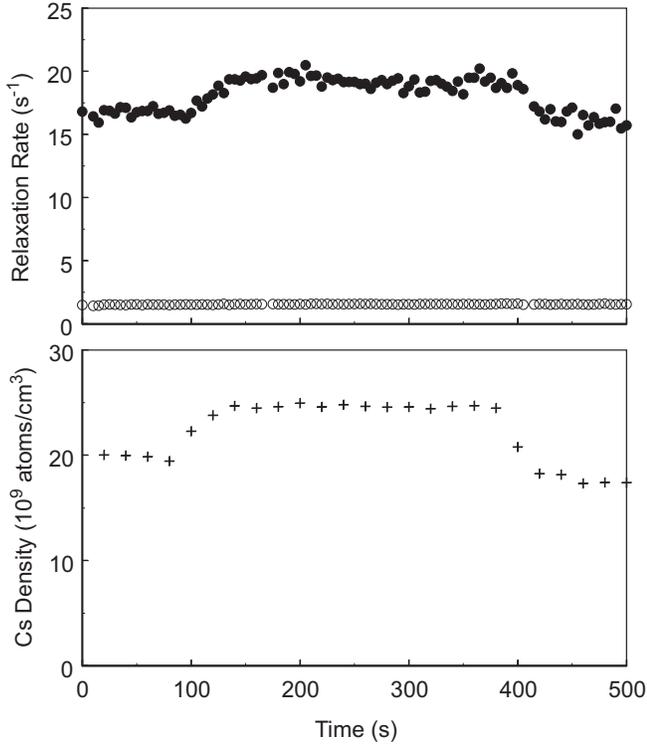}
\caption{Upper plot shows relaxation rates as a function of time
when cell is exposed to off-resonant light that causes desorption
of Cs atoms from the paraffin coating (LIAD, see
Ref.~\cite{ourLIAD}).  Open circles correspond to the slower rate
$\gamma_s$, filled circles correspond to the faster rate
$\gamma_f$ [see Eq.~\eqref{Eq:FittingFunction}]. Lower plot shows
the change in Cs density when the cell is exposed to the desorbing
light. Desorbing light at 514 nm has intensity $\approx 7~{\rm
mW/cm^2}$, and is activated at $t=100~{\rm s}$ and turned off at
$t=380~{\rm s}$. Laser frequency is detuned $\approx 400~{\rm
MHz}$ to the high-frequency wing of of the $F=4 \rightarrow F'$
hyperfine component of the Cs D2 absorption line (where the
amplitudes of the time-dependent signals are large, see
Figs.~\ref{AmpRateSpectrum} and \ref{AmpRateSpectrum}), pump light
power = 4 mW, probe light power = 3 $\mu$W, $|\vec{B}|=10~{\rm
G}$, cell temperature = $20^\circ{\rm{C}}$.}\label{LIADtimedep}
\end{figure}

The relaxation rates before, during, and after exposure of a cell
to desorbing light are shown in Fig~\ref{LIADtimedep}. Cell A
(Table~\ref{Table:CellDimensions}) was fully illuminated using the
Ar$^+$ laser. The results show a correlation between the fast
relaxation rate and the presence of desorbing light, whereas the
slow relaxation rate appears to be unaffected by the presence of
desorbing light.

\begin{figure}
\includegraphics[width=3.35 in]{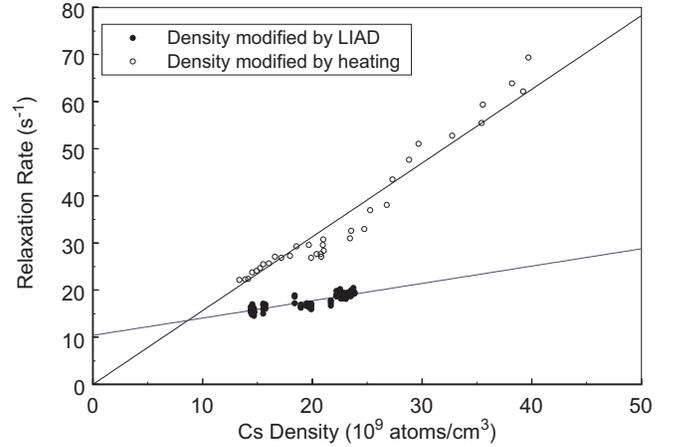}
\caption{Fast rate as a function of Cs density, where in one case
the Cs density is changed by heating the cell, and in the other
case the Cs density is changed by using off-resonant light to
desorb atoms from the paraffin coating (LIAD). The factor of four
difference in the slopes is evidence of extra relaxation induced
when the cell is heated. The slope obtained using LIAD to control
the density is consistent with that expected from Cs-Cs
spin-exchange relaxation.  All data are taken with cell A
(Table~\ref{Table:CellDimensions}). Cell conditions described in
previous figure captions (Fig.~\ref{LIADtimedep} for LIAD and
Fig.~\ref{TempDep} for heating.)}\label{FastRate_LIADvsHeating}
\end{figure}

In Fig.~\ref{FastRate_LIADvsHeating}, the fast relaxation rate is
shown both when LIAD is used to change Cs density $n$ and when the
cell is heated. As discussed previously in
Sec.~\ref{SubSec:CellTempDep}, the slope of the fast relaxation
rate with respect to $n$ is a factor of four larger than the value
expected from spin-exchange collisions when $n$ is changed by
heating the cell.  On the other hand, the Cs density dependence of
the fast rate when LIAD is used to change $n$ is consistent with
relaxation due solely to spin-exchange collisions [according to
fits to Eq.~\eqref{Eq:FastandSlowRates}]. Thus it appears that
LIAD does not change the relaxation properties of the wall
coating, and in fact avoids the ``extra'' relaxation due to
electron randomization collisions introduced when the entire cell
is heated.

Note that the fast relaxation rate does not extrapolate to zero
for $n=0$ when LIAD is used to alter $n$, presumably because of
the presence of the ``extra'' relaxation described by the rate
$\gamma_{er}$. Furthermore, one may notice that at every Cs
density, the data for $\gamma_f$ are different for the LIAD
experiment and the heating experiment. This can be understood as a
result of the fact that the temperature of the cell (20$^\circ$C)
in the LIAD experiment is lower than any of the temperatures
during the heating experiment (minimum temperature =
22.4$^\circ$C) -- note that at lower Cs densities ($\approx
10^{10}~{\rm{atoms/cm^3}}$) the extrapolations from the fits
intersect.  This may be an indication of a temperature-dependent,
Cs-density-independent source of spin relaxation \endnote{However,
there is some ambiguity because temperature and Cs density are not
always reproducibly correlated from run-to-run in the
paraffin-coated cells we use -- occasionally it is necessary to
temporarily overheat the stems (for about 5 minutes) relative to
the body of the cells to reach the usual vapor pressure at room
temperature.}.

To confirm that the way in which the cell is illuminated does not change these results,
we performed an experiment in which only a small portion (12\%) of the cell surface was
exposed to desorbing light.  The light intensity was increased (from $\approx~7~{\rm
mW/cm^2}$ to $\approx~28~{\rm mW/cm^2}$) so that the overall change in density was
comparable to the data shown in Fig.~\ref{FastRate_LIADvsHeating}.  We observed a nearly
identical change in the fast relaxation rate.

\begin{figure}
\includegraphics[width=3.35 in]{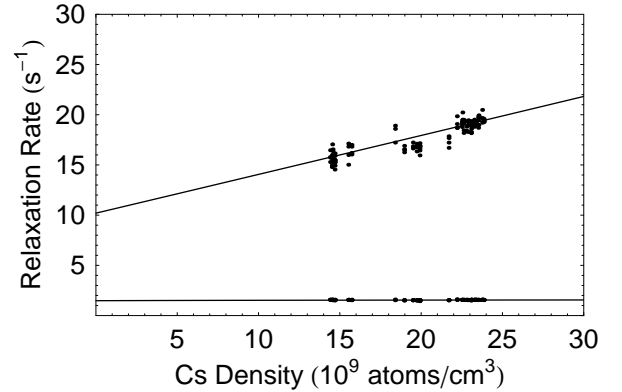}
\caption{Comparison of theoretical prediction of
relaxation-in-the-dark rates based on
Eq.~\eqref{Eq:FastandSlowRates} where for the spin-exchange rate
$\gamma_{se}=n\sigma_{se}v\ts{rel}$ we employ
$\sigma_{se}=2\times10^{-14}~{\rm cm^2}$ determined by previous
measurements \cite{Bouchiat}, and we determine that
$\gamma_{er}=9.0(1)~{\rm s^{-1}}$ and $\gamma_u=1.2(1)~{\rm
s^{-1}}$ from fitting the data.  LIAD is used to change the Cs
vapor density in this case, all data are taken with cell A
(Table~\ref{Table:CellDimensions}), cell conditions described in
the caption of Fig.~\ref{LIADtimedep}.}\label{RID_DataAndTheory}
\end{figure}

Figure \ref{RID_DataAndTheory} compares the density dependences of
the fast and slow relaxation rates predicted by the model
described in Sec.~\ref{SubSec:PrincipleOfRIDOVOR} to the data
obtained when LIAD is used to alter $n$. As predicted, the fast
and slow rates have the dependence given by
Eq.~\eqref{Eq:FastandSlowRates} using the known $n$-dependence of
spin-exchange relaxation in Cs.  Extrapolation to $n=0$, and thus
$\gamma_{se}=0$, yields $\gamma_u$ and $\gamma_{er}$ from
Eqs.~eqref{Eq:RatesAtZeroDensity1} and
\eqref{Eq:RatesAtZeroDensity2}. The analysis clearly shows that
electron-randomization collisions are a dominant source of
relaxation in the cell at room temperature and above, in agreement
with the results of Ref.~\cite{NISTpaper}.

Finally, in an effort to establish the specific location of the
collisions causing electron randomization relaxation, we used LIAD
to investigate the relaxation rates in cell C (the small cell,
Table~\ref{Table:CellDimensions}). This relaxation could be
occurring in collisions with the wall or possibly in collisions
with some gaseous impurity in the cells.  If the relaxation is
occurring in wall collisions, the effect should scale inversely
proportional to the cell radius:
\begin{align}
    \gamma_{er}= \gamma\ts{wall} \sim P \frac{v}{R}
\end{align}
where $\gamma\ts{wall}$ is the electron-randomization relaxation
rate on the wall, $P$ is probability for relaxation in a single
collision with the wall, and $R$ is the effective radius of the
cell (half the mean free path listed in
Table~\ref{Table:CellDimensions}). We used a Monte-Carlo
simulation to determine $R$, assuming a cosine distribution of
atoms reflected from the paraffin surface. Relaxation due to
gaseous impurities should have no dependence on the size of the
cell.

For a reasonable comparison of relaxation in the cells of two
different sizes, much higher light intensity had to be used for
the small cell C in order to obtain a stable and comparable change
in density (this is because of the difference in the ratio of the
cells' surface areas to stem entrance areas, as discussed in
Ref.~\cite{ourLIAD} -- the stem acts as a pump for the excess
vapor density produced in the volume of the cell by LIAD). To
accomplish this, we used the same lens set-up as in the
aforementioned partial illumination experiment and increased the
desorbing light intensity to $\approx~120~{\rm mW/cm^2}$. The
rates $\gamma_{se}$, $\gamma_{er}$, and $\gamma_{u}$ were
determined from the data by fits to our model. For the small cell,
$\gamma_u = 15(1)~{\rm s^{-1}}$, consistent with a larger
reservoir effect due to the lower surface area to stem entrance
area ratio compared to cell A (Table~\ref{Table:CellDimensions}).
The electron randomization rate is $\gamma_{er} = 18(1)~{\rm
s^{-1}}$ for cell C, an increase of 2 times compared to cell A.
This increase is comparable to ratio of effective radii of the
cells, $R_A/R_C \approx 2.2$, consistent with the notion that this
relaxation is on the cell walls \cite{NISTpaper}.

If the relaxation rate $\gamma_{er}$ can be attributed to
electron-randomizing wall collisions, why does heating the cell
increase $\gamma_{er}$ while LIAD seems to not affect it?  The
most straightforward explanation would be that $\gamma_{er}$
depends on the wall coating temperature, but this would contradict
previous studies which show that relaxation properties of the
coating actually improve at higher temperatures (up to about
$60^\circ$C) \cite{Vanier}. In the study described in
Ref.~\cite{Vanier}, as previously discussed, data was extrapolated
to zero Rb density to isolate effects dependent on the coating
temperature from effects dependent on the Rb density, and electron
randomization collisions were not found to dominate relaxation.
This leads us to the conclusion that $\gamma_{er}$ is caused by
some alkali-atom induced modification of the coating surface. But
why does increasing the vapor density with LIAD not cause a
similar modification of the coating surface?

Alkali-atom-induced modification of the paraffin coating is known
to occur.  There is considerable experimental evidence (see
Ref.~\cite{ourLIAD} and references therein) suggesting that during
the cell preparation procedure alkali atoms react with
paramagnetic impurities in the paraffin-coating, thereby
eliminating paramagnetic sites from the paraffin surface and
subsequently improving the relaxation properties of the coating.
It may be that when the cell coating is heated, some fraction of
Cs atoms previously bonded to the paramagnetic impurity sites are
released and the sites become active again, increasing the
probability of electron-randomizing wall collisions. Conversely,
when LIAD is used to desorb Cs atoms from the coating, Cs atoms
not associated with the paramagnetic relaxation sites are
released.  However, this hypothesis would again contradict
previous work showing a decrease in relaxation due to wall
collisions at higher temperatures \cite{Vanier}.

One possible explanation that seems consistent with both the
present work and previous studies of relaxation properties of
paraffin-coated cells
\cite{Robinson,Bouchiat,Alexandrov1,Alexandrov2,NISTpaper,BouchiatPhD,Bouchiat,Gibbs,Liberman,BalabasRID,Vanier,AleksandrovK}
would involve a alkali-atom induced modification of the surface
that takes a relatively long period of time to occur.  In this
case, increasing the alkali density for a short time using LIAD is
insufficient to alter the coating properties.  An increase of
density sustained over a longer period of time, as in the case of
heating the cell, may allow sufficient time for the surface
modification.

Clearly, further experimentation is required to resolve this
issue.

\section{Conclusion}

We have studied the relaxation of optically pumped ground-state
atomic polarization in paraffin-coated cesium vapor cells by
measuring ``relaxation in the dark'' using optical rotation of a
narrow-band, low intensity probe beam.  The approach employed in
the present work enabled a clear distinction of two different
relaxation rates.  A model of relaxation processes in the cell is
presented that relates the measured relaxation rates to three
different physical relaxation mechanisms in the cell: (1)
spin-exchange collisions between Cs atoms, (2)
electron-randomization collisions with, for example, the cell
wall, and (3) a process that relaxes all atomic polarization
moments at the same rate, for example due to exchange of atoms
between the metal sample in the stem of the cell and the vapor
phase in the volume of the cell -- the ``reservoir effect''
\cite{Bouchiat}.

The relaxation rates were studied as a function of pump and probe
light power and detuning, magnetic field, and cell temperature.
These studies confirmed that the assumptions made in our model
were reasonable and that the model could be used to extract
information about relaxation processes in the cell. The change in
relaxation rates when the cell temperature was increased greatly
exceeded that expected from an increased rate of Cs spin-exchange
collisions -- indicating that there existed some additional,
cell-temperature-dependent relaxation process.

Relaxation rates were also studied when the cells were exposed to
off-resonant light that caused desorption of Cs atoms from the
paraffin coating (Light-Induced Atomic Desorption --- LIAD
\cite{ourLIAD}).  When the Cs vapor density increased in the cell
due to LIAD, the spin relaxation rates changed as expected
assuming only the spin-exchange rate increased. This technique
enabled us to unambiguously separate the contributions of the
three physical relaxation processes described in our model. It was
determined that spin relaxation in the cells was dominated by
electron-randomization collisions. Comparison of rates in
differently sized cells indicated that collisions with the cell
walls were the source of the electron randomization (see also
Ref.~\cite{NISTpaper}). Furthermore, we found no evidence of a
change in the relaxation properties of the coating during LIAD.

This study has demonstrated that LIAD is a promising tool for the
control of alkali vapor density in paraffin-coated cells, since it
does not change spin-relaxation properties of the coating.
Additionally, the use of LIAD in conjunction with
relaxation-in-the-dark measurements enabled some basic
characterization of the relaxation processes in paraffin-coated
cells.

LIAD may become a useful technique in the ongoing work aimed at
developing highly miniaturized atomic frequency references
\cite{SmallClocks1,SmallClocks2} and magnetometers
\cite{SmallMags}. These devices will likely take advantage of
miniature atomic vapor cells with physical dimensions on the order
of 1 mm or smaller \cite{SmallCells}. Because of the larger
surface-to-volume ratio, atoms confined in such a small cell spend
a larger fraction of their time interacting with the cell wall
than they would in a larger cell -- so understanding relaxation
due to wall collisions becomes even more important.  Additionally,
in order to increase the signal in such miniaturized cells, an
efficient method to increase the atomic density is required.
Compared to changing vapor densities by heating the cells, our
work shows that LIAD may offer significant improvement in
spin-relaxation times.

Because of the importance of anti-relaxation coated cells in many
areas of research, it is crucial to develop a complete
understanding of the spin-relaxation processes in coated cells. In
the future, we plan to apply modern surface science techniques to
this interesting and important question, potentially leading to a
design of a new generation of anti-relaxation coatings with even
better spin-relaxation properties.

\section{Acknowledgments}

The authors would like to sincerely thank Damon English for Monte
Carlo simulations of the atomic trajectories in the
parrafin-coated cells, and A. I. Okunevich for useful comments and
suggestions on the manuscript.

This research was supported by the Office of Naval Research (grants N00014-97-1-0214 and
SBIR), the Russian Foundation for Basic Research (grant 03-02-17509 RFBR), the National
Science Foundation, a Cal-Space Mini-grant, Lawrence Berkeley National Laboratory's
Nuclear Science Division, and the University of California at Berkeley's Undergraduate
Research Apprenticeship Program.

\

\end{document}